\documentstyle[11pt,newpasp,twoside]{article}

\begin{document}

\title{Analytic Approaches to Understanding the
Distribution and Evolution of Intracluster Gas}

\author{Xiang-Ping Wu}
\affil{National Astronomical Observatories, Chinese Academy
                of Sciences, Beijing 100012, China}

\begin{abstract}
I will briefly review various analytic approaches to understanding the
contents and properties of the hot X-ray emitting gas contained
in clusters and groups. Special emphases are given to the following
three issues: (1)Reconstruction of the gas
distribution in groups and clusters from a (joint) analysis of
X-ray, SZ and gravitational lensing observations;
(2)Test of the analytic density profiles of dark halos suggested by
numerical simulations and empirical models with current X-ray data;
And (3)the effects of  preheating and radiative cooling
on the X-ray properties of groups and clusters.
\end{abstract}

\section{Introduction}

Groups and clusters serve as a reservoir of baryons in the present-day
universe. They exist in the form of hot plasma with temperature
close to the virial temperature ($10^6$-$10^8$ K) of the underlying
gravitational potential wells as a result
of gravitationally-driven shocks and adiabatic compression.
Since the discovery of diffuse X-ray emission associated with
clusters of galaxies, many efforts have been made towards
exploring the distribution and evolution of the hot gas in clusters.
This is significant not only for our understanding
of the matter composition and dynamical properties of clusters
but also for test of various theories of structure formation.
In recent years, a combination of multi-wavelength observations
(e.g. optical, X-ray, radio, gravitational lensing, etc.),
theoretical analysis and hydrodynamical simulations has greatly improved
our knowledge of the intragroup/intracluster gas.
In this review I will concentrate on the analytic approaches to
determining and modeling the distribution and evolution of the
diffuse X-ray emitting gas in groups and clusters.

\section{Observational approaches}

\subsection{X-ray observations}

In principle one is able to analytically derive the gas distribution
of clusters by direct inversion of the X-ray surface brightness
profile $S_x$.
Under the assumptions of spherical symmetry and  thermal Bremsstrahlung,
we have (e.g. Cowie,  Henriksen \& Mushotzky 1987)
\begin{equation}
n_e^2(r)\Lambda(T)=\frac{4(1+z)^4}{r} \frac{d}{dr}
           \int_r^{\infty}S_x(R)
           \frac{RdR}{\sqrt{R^2-r^2}},
\end{equation}
where the emissivity $\Lambda$ is roughly proportional to
$T^{1/2}$ for clusters. The electron density profile $n_e(r)$ can thus
be obtained in combination with high resolution spectral observation.
In the case of isothermality and for conventional $\beta$
model $S_x\propto(1+r^2/r_c^2)^{-3\beta+1/2}$,
The above equation reduces to the well-known functional form of
\begin{equation}
n_e(r)=n_0\left(1+\frac{r^2}{r_c^2}\right)^{-3\beta/2}.
\end{equation}
Even for the double $\beta$ model suggested by some authors
(e.g. Ikebe et al. 1996; 1999; Xu et al. 1998; Mohr,
Mathiesen \& Evrard 1999; Xue \& Wu 2000c; Ettori 2000) aiming at
a better description of the central excess emission associated with
cooling flows,  an analytical form of gas density can also be achieved.
Yet, the accuracy of the derived gas density
profile may suffer from  uncertainties not only in the determination of
temperature profile but also in the extrapolation of the available X-ray
data to the outer-skirts of clusters. Moreover,
the inversion of the X-ray surface brightness profiles becomes complicated
if the hot gas in clusters actually has an asymmetrical distribution.
Another technique, which has been widely used in the analysis
of X-ray clusters,
is to deproject the X-ray surface brightness by
specifying {\it a priori} the gravitational potential of a cluster
and demanding additionally hydrostatic equilibrium (Fabian et al. 1980, 1981;
White, Jone \& Forman 1997). Strictly speaking, such a method is
not a purely observational approach to the reconstruction of
the distribution of intracluster gas.

\subsection{Combination of X-ray and optical observations}

The pioneering effort to derive the empirical $\beta$ model of
gas density in clusters can be traced back to 1976 when
Cavaliere \& Fusco-Femiano (1976, 1978) combined the Jeans
equation for galaxies
\begin{equation}
\frac{1}{n_{\rm gal}(r)}\frac{d[n_{\rm gal}(r)\sigma_r^2(r)]}{dr}
=-\frac{d\Phi(r)}{dr}
\end{equation}
with the hydrostatic equation for gas
\begin{equation}
\frac{1}{\rho_{\rm gas}}\frac{d[n_{\rm gas}(r)kT(r)]}{dr}=
-\frac{d\Phi(r)}{dr},
\end{equation}
and reached
\begin{equation}
n_{\rm gas}(r)\propto n_{\rm gal}^{\beta_{\rm spec}}(r)
   \propto \left(1+\frac{r^2}{r_c^2}\right)^{-3\beta_{\rm spec}/2},
\end{equation}
in which an isotropic, constant velocity dispersion of galaxies and
a constant gas temperature have been adopted, and the last equality in
the above equation assumed a simplified King model for the radial
variation of galaxy number density,
$n_{\rm gal}\propto(1+r^2/r_c^2)^{-3/2}$. This does reproduce
the functional form of the empirical $\beta$ model except the power index:
Direct fitting of the X-ray surface brightness of clusters gives
a mean value of $\langle\beta\rangle\approx0.6-0.7$,
while the average spectral parameter [$\beta=\sigma^2/(kT/\mu m_p)$]
seems to be slightly larger, $\langle\beta_{\rm spec}\rangle\approx1$.
This so-called $\beta$ discrepancy has attracted the attention of many
investigators (e.g. Bahcall \& Lubin 1994; Gerbal, Durret
\& Lachi\`eze-Rey 1994;  Bird, Mushotzky \& Metzler 1995; Wu, Fang
\& Xu 1998; Xue \& Wu 2000c; etc.), and several possible reasons
including observational selection effect have been extensively
discussed in the literature.

\subsection{Combination of X-ray and S-Z measurements}

Silk \& White (1978) were the first to suggest the utilization of
the nonparametric reconstruction of the radial profiles of density
[$n_e(r)$] and temperature [$T(r)$] of intracluster gas
by combining the X-ray and SZ measurements. The basic idea is to
inverse the observed X-ray and SZ temperature surface brightness profiles of
clusters:
\begin{eqnarray}
n_e^2(r)T^{1/2}(r)&\propto& \frac{1}{r} \frac{d}{dr}
           \int_r^{\infty}S_{\rm X}(R)\frac{RdR}{\sqrt{R^2-r^2}};\\
n_e(r)T(r)        &\propto& \frac{1}{r} \frac{d}{dr}
           \int_r^{\infty}S_{\rm SZ}(R)\frac{RdR}{\sqrt{R^2-r^2}},
\end{eqnarray}
where $S_{\rm SZ}(R)$ is the SZ surface brightness across clusters.
And a simple combination of the two functions gives straightforwardly
the gas density $n_e(r)$ and temperature $T(r)$.
This method requires no assumptions about the dynamical
properties of clusters and the equation of state for intracluster
gas, and can therefore be regarded as an ideal and ultimate tool to probe
the gas distribution in clusters under spherical approximation.

Despite its elegant mathematical treatment further developed by
Yoshikawa \& Suto (1999) based on some theoretical models,
the pioneering suggestion of Silk \& White (1978)
has not yet been put into practice.
This has been primarily limited by the instrumental sensitivity and
resolution of detecting the temperature variations of the
cosmic microwave background (CMB) behind clusters.
So far, marginal detections of the radial SZ temperature distributions
have been reported only for a few clusters (see Birkinshaw 1999),
which can hardly be used for the purpose of reconstructing
the gas temperature profiles because of the sparse data points
and the large associated uncertainties. A preliminary analysis of
the central gas temperatures of 31 clusters provided by the
two methods has been made,  based on the $\beta$ model for the
gas density profiles (Zhang \& Wu 2000). It turns out that there is
a good agreement between these two independent temperature estimates.

\subsection{Combination of X-ray, SZ and weak lensing measurements}

Weak gravitational lensing technique provides a robust
constraint on the shape of the gravitational potential of clusters.
With this additional observation, one will be able to reconstruct
the intracluster gas distribution even if it is not spherical
(e.g. Cooray 1998; Zaroubi et al. 2001; Fox \& Pen 2002).

The X-ray surface brightness in a given energy band is simply
\begin{equation}
S_{\rm x}(r)=\frac{1}{2\pi(1+z)^3} \int_r^{\infty}
            n_e^2\Lambda(T,n_e)d\ell.
\end{equation}
The SZ temperature variation caused by a cluster is
\begin{equation}
\frac{\Delta T_{\rm SZ}}{T_{\rm CMB}}=g(x)\;\frac{\sigma_T}{m_ec^2}
         \int n_e\;kT_e\; d\ell,
\end{equation}
where $g(x)$ denotes the spectral dependence and is approximately
equal to $-2$ in the Rayleigh-Jeans regime,
$x=h_{\rm p}\nu/kT_{\rm CMB}$, and $T_{\rm CMB}=2.728$ K
is the temperature of CMB. Under the assumption of spherical
symmetry, the inversion of these two equations yields
the radial profiles of gas density $n_e$ and temperature $T(r)$,
as has been discussed above. However, if the gas distribution deviates
from spherical symmetry, additional constraints must be postulated.
The equation of hydrostatic equilibrium is certainly the first choice:
\begin{equation}
\nabla p=-\rho_{\rm gas}\nabla\phi.
\end{equation}
Here $\phi$ represents the gravitational potential of the cluster, which
should be either determined by other independent observations or specified
by numerical simulations or theoretical models. Weak gravitational lensing
permits a direct reconstruction of the projected cluster mass
$\Sigma(\mbox{\boldmath $\theta$})$, which is related to the two-dimensional
potential $\psi$ through
\begin{equation}
\bigtriangledown^2\psi=\frac{2\Sigma(\mbox{\boldmath $\theta$})}{\Sigma_c}
                     =2\kappa,
\end{equation}
where $\Sigma_{\rm crit}=(c^2/4\pi G)(D_s/D_dD_{ds})$ is the critical
surface density, and $\kappa$ is called convergence. Finally,
\begin{equation}
\psi(\mbox{\boldmath $\theta$})=
\frac{1}{2\pi G\Sigma_{\rm crit}}\int \phi d\ell.
\end{equation}
With this additional constraint from gravitational lensing, one can
now reconstruct the gas distribution  if the cluster possesses
an axial symmetry. Strictly speaking, such a combined analysis is not
a purely observational method because it demands the condition of
hydrostatic equilibrium for the intracluster gas.

\section{Tests of analytic density profiles for dark halos}

High-resolution cosmological simulations suggest that the dark matter
density profile is well described by a self-similar, two-parameter
functional form (e.g. Navarro, Frenk \& White 1997):
\begin{equation}
\rho_{\rm DM}(r)=\frac{\rho_{\rm crit}\delta_c}
       {(r/r_s)^{\alpha}(1+r/r_s)^{3-\alpha}}.
\end{equation}
The parameters $\delta_c$ and $r_s$ set the scale of the mass distribution
in density and length, and $\alpha$ is the logarithmic power-law slope
near the center, with $\alpha=1-1.5$ (Navarro et al. 1997; Moore et al.
1998).  Other empirical density profiles advocated in the literature
include Hernquist profile $\rho_{\rm DM}(r)=(M/2\pi a)/[(r/a)(1+r/a)^3]$ and
Burkert profile $\rho_{\rm DM}(r)=\rho_0/[(1+r/r_0)(1+r^2/r_0^2)]$.
Now our strategy is
to find the distribution of the intracluster gas if it is in
hydrostatic equilibrium with the underlying gravitational potential of
dark matter defined by one of the above analytical forms:
\begin{equation}
\frac{1}{\mu m_p n_{\rm gas}(r)}\frac{d[n_{\rm gas}(r)kT(r)]}{dr}=
 -\frac{GM_{\rm tot}(r)}{r^2}.
\end{equation}
This enables one to work out the gas distribution once the equation of state
is specified. Eventually, an examination of the predicted
and observed X-ray surface brightness distributions of clusters will provide
a critical test for the proposed dark matter profiles.  In the case of
isothermality for the gas and an NFW profile for the dark matter distribution,
one has $n_{\rm gas}(r)\propto (1+r/r_s)^{b/(r/r_s)}$ where
$b=4\pi G\mu m_p \rho_sr_s^2/kT$ (Makino, Sasaki \& Suto 1998).
This functional form can be approximately fitted by the $\beta$ model
except the small core sizes as compared with X-ray observations of clusters.
Inclusion of the temperature gradient parameterized by the polytropic
form seems to provide a more plausible prediction
(Suto, Sasaki \& Makino 1998) but a direct comparison with observations
is hampered by the poorly constrained temperature profiles at present.
Alternatively, one can also assume that the gas distribution is
self-similar, i.e., $n_{\rm gas}=(f_b/\mu m_p)\rho_{\rm DM}(r)$
where $f_b$ is the universal baryon fraction.
This is justified by a number of hydrostatic simulations at least outside
the central cores of clusters (e.g. Bryan \& Norman 1998;
Eke, Navarro \& Frenk 1998; Pearce et al. 2000; etc.).
As a result, we can predict the radial profile of gas temperature
\begin{equation}
kT(r)=\left(\frac{\mu m_pG}{\rho_{\rm DM}(r)}\right)
     \int_r^{\infty}\rho_{\rm DM}(r)\frac{M(r)}{r^2}dr,
\end{equation}
in which we have neglected the self-gravity of the intracluster gas and
set $n_{\rm gas}(r)kT(r)$ to approach $0$ when $r\longrightarrow\infty$.
One can equally use the central temperature $T(0)$ as the boundary condition
and/or the polytropic state of equation to get more general form of
the temperature profile (e.g. Komatsu \& Seljak 2001).
Another possible way to derive the temperature profile is to combine
the high-resolution surface brightness measurement $S_x$ with the equation of
hydrostatic equilibrium for a given dark matter distribution such as the
NFW profile (Wu \& Chiueh 2000; Xue \& Wu 2000a). Yet,  high quality,
spatially-resolved spectral measurements will be needed
to test these predictions.

The striking similarity between the predicted X-ray surface brightness
of clusters by NFW profile in the case of isothermality
and the conventional $\beta$ model has stimulated several authors
to determine the two free parameters in the NFW profile through a direct
fitting of the observed data of clusters to theoretical prediction
(e.g. Makino \& Asano 1999; Ettori \& Fabian 1999; Wu \& Xue 2000a,b;
Xu, Jin \& Wu 2001; etc.).
Although the best fit values of $r_s$ and $\delta_c$ have been given
for a few tens of X-ray clusters, great caution must be exercised if
these data are used for cosmological purposes (Wu \& Xue 2000a;
Sato et al. 2000; Cheng \& Wu 2001). Note that the fitting process is
sensitive to the data points at the central regions
of clusters, in which the presence of central cooling flows in most of clusters
may lead to large uncertainties in the determination of $r_s$
($r_s\approx r_c/3$).

In a word, current measurements of the X-ray surface brightness of clusters
are still insufficient to distinguish various proposed analytic models for
dark matter distribution in clusters except the Burkert profile.
This last empirical profile is based on the fitting of the rotation
curves of the dark matter dominated dwarf galaxies (Burkert 1995).
Such a model resembles an isothermal profile in the inner region with
a constant core $r_0$, while in the outer region the mass profile
diverges logarithmically with $r$, in agreement with the generalized
NFW profile. The model has recently received considerable attention
because of the existence of the soft inner matter
distributions of the dark halos claimed by some observations
(e.g. Flores \& Primack 1994;
de Blok \& McGaugh 1997; Tyson \& Kochanski \& dell'Antonio 1998; etc.).
In despite of its success on galactic scales, the Burkert profile
predicts too large core radii of dark matter profiles for clusters
to be reconciled with the measurements of strong gravitational lensing
(Wu \& Xue 2000b; Xue \& Wu 2001). Meanwhile, its predicted X-ray
surface brightness of clusters fails in the fitting of the observed
data (Xu et al. 2001).

\section{Physical motivations}

While gravity plays a dominant role in the overall distribution and evolution
of the hot gas in groups and clusters, the gas also suffers from
the influence of nongravitational effects such as
radiative cooling, nongravitational heating, nonthermal
pressure, etc. In low mass groups and the central regions of clusters
the nongravitational effects could even govern the behavior of the gas.
Indeed, it has been realized that some peculiar
X-ray features of groups and clusters are associated with
nongravitational processes, although the essential physics
still remains a subject of intense debate.
In the following discussion,
we will focus on the impacts of preheating and radiation cooling
on the X-ray properties of groups and clusters.

\subsection{Observational evidence}

There is growing observational evidence for the presence of nongravitational
heating and/or radiative cooling in groups and clusters which suppresses
the X-ray emission of the gas heated by purely gravitational shocks
and adiabatic compression.
A summary of the observational facts is given below:

$\bullet$
The steepening of the X-ray luminosity - temperature relation of groups
and clusters (David et al. 1993; Wu, Xue \& Fang 1999; Xue \& Wu 2000b
and references therein);

$\bullet$
The existence of the entropy floor in the central cores of groups
and clusters (Ponman, Cannon \& Navarro 1999);

$\bullet$
The flattening of the X-ray surface brightness profiles of poor clusters
and groups (Ponman et al. 1999);

$\bullet$
The upper limit on the X-ray background from the diffuse gas bound
in groups and clusters (Pen 1999; Phillips, Ostriker \& Cen 2001;
Wu \& Xue 2001 and references therein);

$\bullet$
The scale-dependence of the gas and stellar mass fractions and
the mass-to-light ratios of groups and clusters (Mohr et al. 1999;
Bryan 2000; Bahcall \& Comerford, 2002; Wu \& Xue 2002b;
Girardi  et al. 2002; etc.);

$\bullet$
A monotonic increase in the gas mass fraction of clusters with cluster
radii (White \& Fabian 1995; Ettori \& Fabian 1999; Markevitch et al.
1999;  Wu \& Xue 2000c; etc.);

$\bullet$
The failure of detection of X-ray halos surrounding spiral galaxies
(Benson et al. 2000).

These independent observations as a whole suggest that there exists
some nongravitational mechanism which removes the content of hot gas
from groups and clusters, and the effect is more significant
in the central regions and low-mass systems. As a consequence,
late-type galaxies contains almost no hot gas at present.
The prevailing physical scenarios proposed
thus far are preheating and radiative
cooling. The former invokes an extra energy source to preheat the gas
to a certain entropy floor so that the heated gas cannot be compressed by
the shallower gravitational potential wells of low-mass systems like
poor clusters and groups, while the latter requires a considerably
high efficiency of cooling to convert hot gas into cooled material
(i.e. stars) inside the systems.  Both processes result in a decrement
of the hot gas in groups and clusters, and have been shown
to explain equally well the observed X-ray properties of groups and clusters
(e.g. Voit et al. 2002; Muanwong et al. 2002; Borgani et al. 2002).

\subsection{Preheating}

The preheating scenario was proposed a decade ago by several
authors (David, Forman \& Jones 1991;
Evrard \& Henry 1991; Kaiser 1991;
White 1991): If the gas was preheated before falling into the
gravitational potentials of groups and clusters dominated by
dark matter, the entropy of the gas would be raised up to a
certain level so that the gas became harder to compress.
The resulting entropy floor breaks the self-similarity between
dark matter and gas, and the flattened radial distribution of
the hot gas developed. For low-mass systems like groups,
the gas may even extend well beyond the viral radii.
This leads to a significant decrease of the gas density
and therefore, the X-ray emission.
An energy budget of 0.4 - 3 keV per gas particle, depending
on the epoch and environment of preheating, is required to
reproduce the observed entropy floor and $L_x$-$T$ relation.
There are two potential energy sources of heating
which have been explored extensively in the literature:
SN explosions and AGN activity. However, a number of recent
studies have found that an unreasonably high efficiency of
energy injection into the intragroup/intracluster medium from
supernovae should be required in order to bring the gas to the
energy level seen in the X-ray luminosity and entropy
distributions of groups and clusters (Wu, Fabian \& Nulsen 1998,
2000; Valageas \& Silk 1999; Tozzi 2001; etc.).
Although the energy supply by AGN activity is sufficiently
large (Valageas \& Silk 1999; Nath \& Roychowdhury 2002),
it still remains unclear how the kinetic energy of the AGN 
activity is deposited onto the baryons.

A phenomenological treatment of the problem is to simply put
the energy budget problem aside, and work directly with 
the observed entropy distribution of intragroup/intracluster gas.
Together with some physically motivated, analytic models and 
empirical relations revealed by hydrodynamical simulations,
one may be able to explain/reproduce the gross observational properties 
of the intragroup/intracluster gas.
Many investigations of these issues have been made over past years.
The shock model suggested by Cavaliere, Menci \&
Tozzi (1997; 1998) has successfully reproduced the
steepening of the $L_x$-$T$ relation and the flat cores of
gas distribution
in the central regions of groups and clusters. The model attempts to
link the preheated infalling gas with the virialized gas
of a dark halo at the boundary through the Rankine-Hugoniot equation.
The new distribution of the gas can be obtained by numerically solving
the equation of hydrostatic equilibrium under the boundary
condition. However, one must reply on Monte Carlo techniques   
in order trace the hierarchical merging history of dark halos.

Balogh, Babul \& Patton (1999) and Babul et al. (2002) developed
a purely analytic model which may help to highlight the essential
physics behind the preheating process.
First of all, the distribution and evolution of the dark matter
component is assumed to be unaffected by preheating and described by
the analytic profile suggested by numerical simulations
(e.g. NFW profile). Moreover, preheating adds an initial,
universal excess entropy $S_0$ to the gas. For low-mass halos,
the maximum velocity of the infalling flow of the preheated gas
induced by gravity
is only subsonic,  and  the accretion shocks by gravitational
collapse thus become less important. In this case, the isentropically
accreted gas with the state of equation $P=K_0\rho_{\rm gas}^{5/3}$ is
completely determined by the equation of hydrostatic equilibrium:
\begin{equation}
\rho_{\rm gas}(r)=\rho_{\rm gas}(r_h)
                 \left[1+\frac{2}{5}\frac{G}{K_0\rho^{2/3}_{\rm gas}(r_h)}
                   \int_r^{r_h}\frac{M(r^{\prime})}{{r^{\prime}}^2}
                        dr^{\prime}\right]^{3/2},
\end{equation}
and the temperature is $kT(r)=\mu m_pK_0\rho_{\rm gas}(r)$, where
$r_h$ is the size of the halo at the epoch of observation.
For massive halos, it assumes that an isentropic core of mass
$M_{\rm gas}(r_c)=(\Omega_b/\Omega_M)M_{\rm isen}$ within radius $r_c$ would
form first. The rest gas will be heated by shocks as the halo grows
more massive. One can use a generalized equation of state to
describe the gas inside and outside the core: $P=K\rho_{\rm gas}^{5/3}$,
where $K=K_0$ for $r\leq r_c$, and $K=K_0(r/r_c)^{\alpha}$ for $r>r_c$.
This last expression is taken from an empirical formula found by
numerical simulations (Lewis et al. 2000). Again, solving the
equation of hydrostatic equilibrium yields
\begin{equation}
\rho_{\rm gas}(r)=\rho_{\rm gas}(r_c)
                 \left[1-\frac{2}{5}\frac{G}{K_0\rho^{2/3}_{\rm gas}(r_c)}
                   \int_{r_c}^{r}\frac{M(r^{\prime})}
                        {{r^{\prime}}^{2+3\alpha/5}}
                        dr^{\prime}\right]^{3/2}.
\end{equation}
Finally, the key point is to specify the turnover
mass ($M_{\rm isen}$) which separates the isentropic accretion from
the accretion shock. This can be achieved by setting the total gas
mass fraction evaluated by the adiabatic Bondi accretion rate
$\dot{M}_{\rm Bondi}$ to equal the universal value:
$M_{\rm gas}/M_h=\dot{M}_{\rm Bondi}t_H(z)/M_h=\Omega_b/\Omega_M$,
in which $t_H$ is the Hubble time.  Of course, in order to properly
normalize the gas density $\rho_{\rm gas}(r_h)$ or $\rho_{\rm gas}(r_c)$,
the universal baryon fraction within $r_h$ or $r_c$ should be assumed.
This may be a major shortcoming for the model since 
the observed gas mass fractions $M_{\rm gas}/M_h$ of clusters seem 
to increase with temperature (e.g. Mohr et al. 1999).

\subsection{Radiative cooling}

Radiative cooling as a natural process was first considered in the formation
of galaxies by White \& Frenk (1991). The major concern has been whether
radiative cooling of the hot gas is sufficient to
account for the X-ray observed properties of groups and clusters
(Bower et al. 2001; Balogh et al. 2001).  Recent hydrodynamical simulations
and semianalytic models have, nevertheless, shown that radiative cooling
alone turns to be successful in the recovery of the entropy excess
and the steepening of the X-ray luminosity-temperature relations
detected in groups and clusters (Bryan 2000; Pearce et al. 2000;
Muanwong et al. 2001, 2002; Voit \& Bryan 2001; Wu \& Xue 2002a,b;
Borgani et al. 2002; Voit et al. 2002),
which may suggest a simple, unified scheme for the evolution of hot gas and
the formation of stars in the largest virialized systems in the universe.

A simple yet plausible approach to evaluating the effect of
radiative cooling is as follows:
We assume that in the absence of cooling the gas has the same
distribution as the dark matter in groups and clusters which
is described, for example, by the universal density profile.
The electron number density of the hot gas thus follows
$n_{\rm e}(r)\propto f_{\rm b}\rho_{\rm NFW}(r)$. We assign an X-ray
temperature to each halo in terms of cosmic virial theorem
$M\propto T^{3/2}$. The hot intragroup/intracluster
gas would continuously lose energy due to bremsstrahlung emission.
The decrease in X-ray temperature $T$ is completely governed  by
the conservation of energy,
\begin{equation}
\frac{3}{2}n_{\rm gas}kT=\epsilon(n_{\rm gas},T,Z)t_{\rm cool}
\end{equation}
where $\epsilon$ is the emissivity. This defines the so-called cooling time
$t_{\rm cool}$ and radius $r_{\rm cool}$ within which gas can cool out of
the hot phase. If the cooling time is set to equal the age of
groups/clusters, or approximately the age of the universe, $t_0$,
we will be able to estimate the maximum cooling radii of the systems
by the present time, $r^{\rm m}_{\rm cool}$, and the corresponding critical
gas density, $n(r^{\rm m}_{\rm cool})$.

The gas within the maximum cooling radius $r^{\rm m}_{\rm cool}$
is assumed to convert into stellar objects.
This latter component should also include the contribution of
other possible cooled materials
(e.g. neutral and molecular gas) which may form out of cooling process.
The cooled gas mass within $r^{\rm m}_{\rm cool}$
can be obtained by integrating the gas
profile $n_{\rm gas}(r)$ over volume out to $r^{\rm m}_{\rm cool}$.
Such a simple exercise gives rather a robust estimate of the total
stellar mass of a system, $M_{\rm star}$, as has been shown recently by
Yoshida et al. (2002). As a result, one can
obtain the stellar and gas mass fractions from
$f_{\rm star}=M_{\rm star}/M$ and $f_{\rm gas}=f_{\rm b}-f_{\rm star}$,
respectively. Because the hot gas cools relatively faster in groups than
in clusters due to the difference in their density contrast,
temperature or metallicity,
poor groups experienced higher efficiency of star formation
than rich clusters did. Therefore, a(n) decrease (increase) of stellar (gas)
mass fraction from groups to clusters is expected to occur
naturally. The predicted scale-dependence of
$f_{\rm gas}$ shows a good agreement with observations (Wu \& Xue 2002b).
However, the cool gas mass fraction slightly exceeds the observed stellar
mass fractions estimated within virial radii by Roussel, Sadat \&
Blanchard (2000), implying that some of the cooling gas may exit in
other forms such as cold clouds. The failure of detecting the cool
gas in clusters might construct a challenge to the cooling model
(e.g. Miller, Bregman \& Knezek 2002).
Nevertheless, within the framework of radiative cooling,
the mass-to-light should increase from groups to clusters because
$M/L=(M/M_{\rm star})(M_{\rm star}/L)=\Upsilon/f_{\rm star}$, where
$\Upsilon\equiv M_{\rm star}/L$ measures the efficiency with which
groups and clusters transform cooled material into light. Indeed, by properly
choosing this parameter, one may be able to quantitatively account for
the mildly increasing trend of $M/L$ with scale claimed recently by
Bahcall \& Comerford (2002) and Girardi et al. (2002).

The newly established gas distribution as a result of radiative cooling
can be obtained by combining the
conservation of entropy and the equation of hydrostatic
equilibrium (Bryan 2000; Voit \& Bryan 2001; Wu \& Xue 2002a;
Voit et al. 2002 ). It appears
that both the cores of the gas density profiles and the entropy
floors are created in the centers as a result of galaxy formation
which has consumed the central hot gas by converting it into stars or
other forms of cold materials.
Meanwhile, a monotonically increasing  gas fraction with radius is
expected to occur. Apparently,
the effect is more significant in lower temperature systems than
in higher ones. It has been shown that
the regulated gas distribution in groups and clusters produced by
cooling or star formation resembles the conventional
$\beta$ model in shape, and the increase of
global $f_{\rm gas}$ with mass provides a natural
explanation of the observed radial variation of $f_{\rm gas}$
(Wu \& Xue 2002a; Voit et al. 2002),
Finally, a straightforward computation of
the X-ray emission of the gas inside groups and clusters permits
a recovery of the observed $L_x$-$T$ relation
(Bryan 2000; Muanwong et al. 2001, 2002; Voit \& Bryan 2001;
Wu \& Xue 2002a; Voit et al. 2002; Borgani et al. 2002).

In a word, radiative cooling of the hot intragroup/intracluster gas,
which is based on the well-motivated physical process, may allow
us to resolve some of the puzzles seen in current X-ray observations
of groups and clusters, and it seems that energy feedback from star
formation comes into effect only in the less massive systems of
$M<10^{13}$ $M_{\odot}$. Yet, current analytic and numerical models
of radiative cooling still have some defects including an overcooling
crisis (Balogh et al. 2001). A critical test of the cooling scenario,
in addition to the conventional test using X-ray properties of
groups and clusters, is the upper limit of the unresolved soft X-ray
background.  It is unlikely that current cooling model can account
for the upper limit of the unresolved soft X-ray background from
the diffuse gas of groups and clusters without excess energy from
preheating (Wu, Fabian \& Nulsen 2001). Indeed, recent hydrodynamic
simulations have suggested that neither preheating nor radiative cooling
alone is able to reproduce all the X-ray observations. A combination of 
both may be the right way to resolve the problem (Borgani et al. 2002).

\subsection{A physically unmotivated model}

If we are interested in the cosmological applications
rather than the internal dynamics and structures
of groups and clusters, we may make a crude estimate of
the influence of nongravitational heating or radiative cooling
using a simple but physically unmotivated approach
(e.g. Holder \& Carlstrom 2001; Voit \& Bryan 2001):
We first evaluate the entropy
distribution of the gas $S^0=kT^0/(n^0_{\rm e})^{2/3}$ by assuming that
the gas is dissipationless and satisfies the equation of hydrostatic
equilibrium. Taking the prevailing NFW profile as an approximation of
the underlying dark matter component, i.e.
$n^0_{\rm e}(r)=(f_b/\mu_e m_p)\rho_{\rm NFW}(r)$ where $\mu_e=1.131$,
we can get the radial profile of the X-ray temperature $T^0(r)$.
Now, regardless of the physical mechanism behind the steepening of the
$L_x$-$T$ relation, the consequence of the effect
must result in an entropy floor in the central regions of groups and
clusters with a value  of $S_{\rm floor}\approx100-400$ keV cm$^2$
at the present epoch. A simple approach is
to define the new entropy distribution as the entropy $S^0$ expected from
self-similar evolution plus a constant entropy floor $S_{\rm floor}$:
\begin{equation}
S(r)=\frac{kT(r)}{n_e^{2/3}(r)}=S_{\rm floor}+S^0(r)
\end{equation}
This is equivalent to specifying the equation of state for the gas.
Now, the new gas distribution is obtained by requiring that the gas is in
pressure-supported hydrostatic equilibrium with the underlying gravitational
potential dominated by dark matter:
\begin{equation}
\frac{n_e(r)}{n_e(r_{\rm vir})}=
         \left[\frac{S(r_{\rm vir})}{S(r)}\right]^{3/5}
                 \left[1+\frac{2}{5}\frac{G\mu_e m_p}
       {n_e^{2/3}(r_{\rm vir}) S(r_{\rm vir})^{2/3}}
                 \int_r^{r_{\rm vir}}\frac{M(r^{\prime})}
                       {S(r^{\prime})^{3/5}{r^{\prime}}^2}
                        dr^{\prime}\right]^{3/2}.
\end{equation}
One needs to normalize the above expression by properly choosing
the gas density at the virial radius $r_{\rm vir}$. To a first
approximation, the total gas density at $r_{\rm vir}$ or the total
gas mass within $r_{\rm vir}$ can be set to equal the universal baryon
fraction $\Omega_b/\Omega_M$. A more plausible approach is
to employ the observationally
determined $f_{\rm gas}$-$T$ relation (e.g. Mohr et al. 1999) or
scale-dependence of stellar mass fraction combined with the universal
baryon fraction.  Although quantitatively crude,
this simple analytic model based on the entropy floor detected
in groups and clusters enables us to effectively demonstrate
the more realistic distribution of the hot gas in today's groups
and clusters.  Yet, the model contains no information about
evolutionary effect unless the variation of the entropy floor
$S_{\rm floor}$ with cosmic epoch can also be given.

\section{Summary}

We have entered a new era of
exploration of matter and energy in groups and clusters,
thanks to the high-sensitivity, high-resolution X-ray observations 
incorporated with optical, radio, SZ and gravitational lensing 
measurements.  A joint analysis of these independent measurements
within next decade will undoubtedly allow us to reconstruct 
more precisely the distributions of both baryons and dark matter 
in groups and clusters, which is especially important with respect 
to the impact on the current debate on the density profiles of 
dark halos suggested by numerical simulations and empirical models.
The physical process of the gas in the formation and evolution of
groups and clusters has been another subject of a longstanding
debate. The prevailing preheating and radiative cooling scenarios 
become indistinguishable in the context of   
current observations, numerical simulations or analytic models. 
This may indicate that our understanding of physical processes 
for gas is still incomplete. Clearly caution thus needs to be 
applied in using groups and clusters for cosmological purpose.

\acknowledgments
I wish to thank S. Bowyer for inviting me to this well-organized
meeting and careful reading of this review.
I would also like to thank L. Y. Lo and T. Chiueh for hospitality 
and useful discussion during my visit at the Institute of Astronomy 
and Astrophysics, Academia Sinica, where this review was prepared.
This work was supported by the National Science Foundation of China,
the Ministry of Science and Technology of China, under Grant
No. NKBRSF G19990754, and
the National Science Council of Taiwan, under Grant NSC91-2816-M001-0003-6.

\end{document}